# A dynamic mechanism of Alzheimer based on artificial neural network

Zhi Cheng

Educational school, Guangzhou University, Guangzhou 510006

Abstract: In this paper, we provide another angle to analyze the reasons why Alzheimer Disease exists. We analyze the dynamic mechanism of Alzheimer Disease based on the cognitive model that established from artificial neural network. We can provide some theoretic explanations to Alzheimer Disease through the analyzing of this model.

Key words: Artificial neural network; Alzheimer Disease; Dynamic mechanism

## 0 Introduction

The meaning of dynamic mechanism is to analyze the principles of an object by using the mathematical method. Since it use the stringent mathematical derivation, the relationships among various factors become more logical, and we can also easy to find where the errors happen. The dynamic mechanism analysis methods have been used in many subject areas, such as classical mechanics, economics and so on.

Alzheimer Disease is a serious cognitive function degeneration symptoms. There is long time history to study the Alzheimer Disease, and also have considerable research data accumulation. But there are no breakthrough prevent or active treatment methods have been obtained..

There was long history in the researching of human cognitive process. The early researches concentrate on the philosophy layer. Psychologists began to study the cognitive process based on experiment data, and obtained abundant results.

The development of artificial neural network theory had caused people's attentions on how to study the human cognitive process from the angle of computation. Since the artificial neural network theory can provide more sufficient method to solve some artificial intelligent problems, the artificial neural network theory was only applied in digital control area. Cheng had proposed the cognitive model based on artificial neural network on 2010[1] in order to solve the human's cognitive development problems. The cognitive development model can obtain the results same as Piaget theory of children's cognitive development[2]. So this model can provide the computational support for Piaget theory. On the other hand, this model also predict there are four development phases for the adult's cognitive development[1,3]. We will do the further improving on this model in this paper.

If we use the cognitive development model based on artificial neural model to study some cognitive abnormalities, we can obtain the new understanding of Alzheimer Disease from a very different angle. We hope this paper can provide some hints for those professionals in relational area.

# 1 The computational complexity of neural network

From the views of artificial neural network theory, we can find why the cognitive depth decreasing is due to the decreasing of neural network computational complexity. It can be shown from the figure 4 of article [1]. Although one's cognitive computational complexity has reached the peak at 30 years of age, his cognitive depth is still increasing. However, when one is at 50 years of age, the curve of computational complexity decreases to meet the curve of cognitive depth in figure 4 of article. It means the curve of computational complexity will limit the further increase of cognitive depth after then. Since then a person officially entered the stage of cognitive ability decline.

## 1.1 Computational complexity and cognitive depth

The relationship between neural network computational complexity and cognitive depth can be represent as two formulas below according to article [1].

$$C(t) = \{m \exp[-h \exp(\frac{t}{\tau})]\}^{aN(t)} \quad\quad\quad (1)$$

$$D(t) = kK(t)[E - lN(t)\frac{dN(t)}{dt}] \quad\quad\quad (2)$$

If D(t) > C(t) Then D(t) = C(t)

From formula (1), we can find that the neural network computational complexity is different from the knowledge accumulation process. The computational complexity is restricted by a person's psychological development. When a person aging, the computational complexity will decrease. In order to understand the origins of neural network computational complexity, here we construct a full interconnecting neural network.

From the formula (2), we can find that the cognitive depth curve is increased in the exponential form. However, this is only one of the knowledge increase model. We believe there are many other knowledge increase models. The different knowledge increase model may also have impacts on the cognitive development process.

All of the neurons are interconnected in this network. Therefore, every neuron's impulse can be received by other neurons.

Considering the neurons impulse number is obey the stochastic distribution, the states of a N(t) neurons network which has n neurons exciting are:

$$S_n(t) = C_{N(t)}^n \quad\quad\quad (3)$$

So the total exciting states are:

$$S(t) = \sum_{n=1}^{N(t)} S_n(t) = \sum_{n=1}^{N(t)} C_{N(t)}^n = 2^{N(t)} \quad\quad\quad (4)$$

Formula (4) is the famous binomial theorem. It reflects the states number of a full interconnecting neural network. The states number can also be considered as the computational complexity of a neural network.

If we consider the neural network's function will decrease as aging. We can use exponential curve to describe the decreasing function. Then the formula (4) can be rewrite as:

$$S(t) = (2e^{-ht})^{N(t)} \quad\quad\quad (5)$$

In formula (5), t is the age, and the unit of age is "month". The coefficient h=0.0001/15

We can use formula (2) and (3) in article [1] to solve the N(t). Notice: the neuron's number in article [1] is represented by I(t).

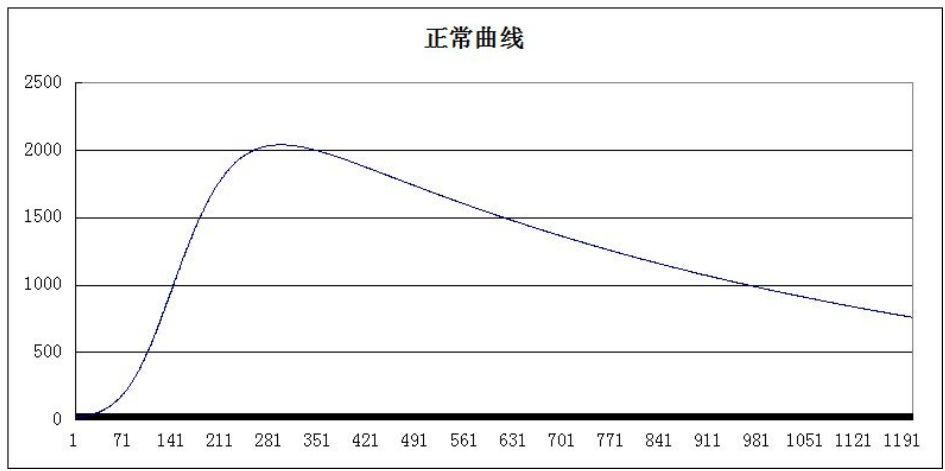

Fig.1 The computational complexity of normal neural network (x-axis: month)

Figure 1 shows the age trends of the normal computational complexity curve. The unit of x-axis is month. We can see the peak of the curve is located in 300th month. It means the cognitive potential of a person will reach his peak in 25 years old. After then, the computational complexity curve will decrease with age. The computational complexity curve will intersect with the cognitive depth curve on 600th month. It means a person's cognitive depth will also reach his peak on 600[th] month according to the results of article [1]. After then, the cognitive depth curve will equal to the computational complexity curve. It will also decrease with age.

Of course, it may be different from each other as every one has their own knowledge accumulation curve. Some one's cognitive depth curve will meet the computational complexity curve more early, others will later.

## 1.2 The coefficients that can affect the computational complexity curve

Therefore, we can see from figure 1 and formula (5), the coefficients that can affect the computational complexity curve are:

1. The growing up of neurons number.
2. The aging of neural network.

# 2 The anomalies of computational complexity curve

It will cause the anomalies of computational complexity curve if the neural network was damaged. There are some conditions can be considered.

## 2.1 The sudden decrease of neurons

After symptoms such as stroke, cerebral hemorrhage, etc., there may be a large number of neurons damaged for the reason of the brain can not get enough oxygen. This will cause reduction of N(t).

The computational complexity curve will change suddenly when neurons decrease 5% in 600[th] month. It is shown in figure 2.

5%

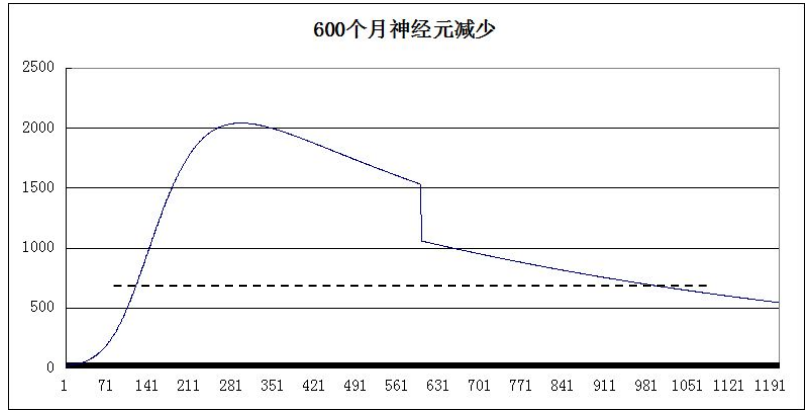

Fig.2 The changes of computational complexity curve caused by the neurons decreasing (x-axis: month)

We can see from figure 2, the curve is still smooth after the sudden reduction of neurons, if there is no other coefficients can lead to the decrease of neurons.

So we can see from figure 2, the 1000[th] month's computational complexity is still equal to the level of 118[th] month. By contrast, the 1000[th] month's normal computational complexity is equal to the level of 138[th] month.

## 2.2 The weaken of neural network function

If we consider the aging process will also cause the exponential decrease of the neural network function. It means the computational complexity curve will decrease more faster. So we can get the computational complexity formula as:

$$C(t) = \{2\exp[-h(e^{\frac{t}{\tau}}-1)]\}^{N(t)} \quad\quad\quad (6)$$

The result of formula (6) is shown in figure 3. The red curve is the normal computational complexity curve that calculated by formula (5). The blue curve is calculated by formula (6). It reflects the neural network weakened more faster.。

From the blue curve, we can find the 1000[th] month computational complexity is equal to the level of 97[th] month.

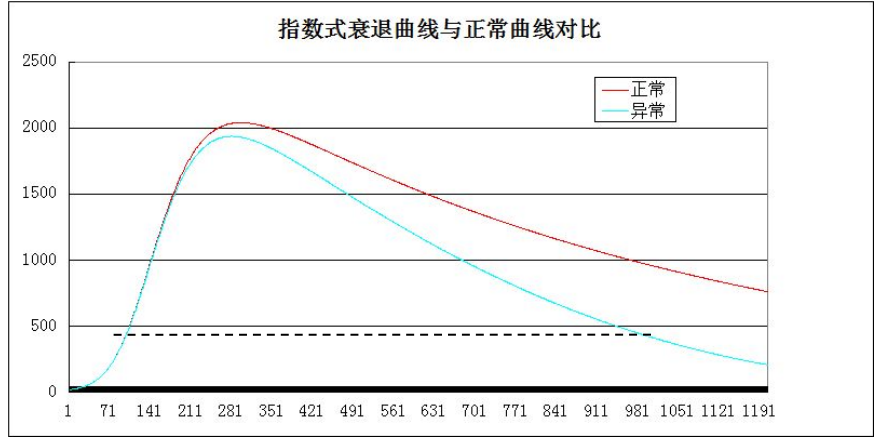

Fig.3 The comparison of normal curve and exponential decreasing curve (x-axis: month)

## 2.3 The coexistence of neurons reduction and neural network weaken

We can combine the figure 2 and figure 3 in this condition. It means that the blue curve in figure 3 will decrease more faster then ever. We can find the $1000^{th}$ month computational complexity is equal to level of $85^{th}$ month in figure 4. Since the computational complexity curve grows up very slowly, it means that the decreasing of computational complexity curve in this condition will damage one's cognitive abilities seriously. It might cause one can not take care of himself.

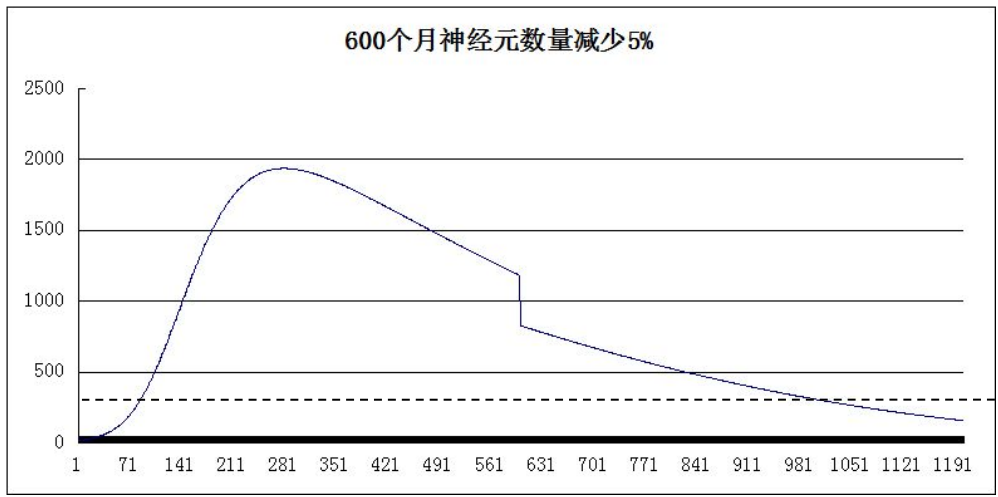

Fig.4 The exponential decrease curve of 5% neurons losing at $600^{th}$ month (x-axis: month)

## 2.4 The sustained decrease of neurons

Figures 2 and 4 are assumed to be a sudden decrease in the number of neurons, and then still be able to maintain a normal rate of change of neurons. If, for some factors that lead to a sustained decrease of neurons, it may lead to computational complexity curve more severe recession. These factors may include the impact of certain drugs and so on. And certain habits may also have some affections, such as tobacco and alcohol.

Figure 5 reflects the neurons is decreased in the rate of 0.05% per month. We can find the computational complexity curve decreased very faster. The cognitive level in $1000^{th}$ month is equal to the level of $45^{th}$ month.

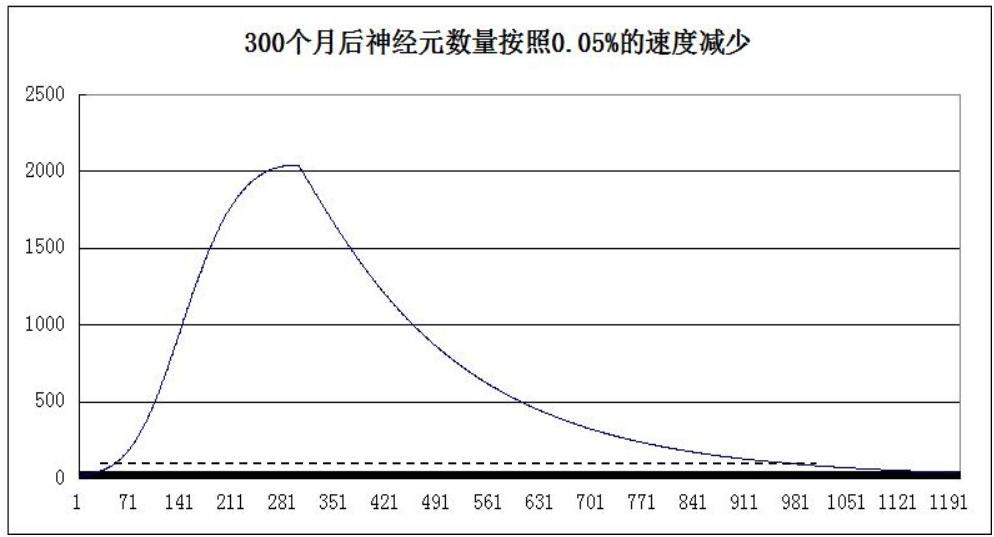

Fig.5 The curve of 0.05% neurons losing velocity after $300^{th}$ month (x-axis: month)

# 3 Analysis

We can find that the human brain neural network is a very stable system from the above analysis. It can maintain the normal physiological functions even if it is damaged by some factors. However, if there are sustained factor that can destroy the neural network, then the brain neural network function will damage seriously.

The adult neural network's computational complexity curve is different from each other. Some one will decrease more faster, it may lead to these adults' cognitive ability appeared just like levels of children. However some others will decrease slowly.

Some conditions may lead to the changes of the computational complexity curve. These conditions include the sudden decrease of neurons; the weaken of neural network function; the coexistence of neurons reduction and neural network weaken; the sustained decrease of neurons.

The first three conditions may cause the computational complexity curve decreasing more faster. However, these conditions just decrease the cognitive abilities in some extents. The adults can still take care of themselves.

The last conditions introduce the sustained damage factor. We can see from the computational results that the computational complexity curve is decreased very faster. The results depend on the decreasing rate sensitively. If the speed of sustained neurons decrease is relatively small, the impact on computational complexity curve is not great. However if the speed is very large, the curve will quickly fall to the limit. Only when we can control the decrease speed to the suitable level, the computational complexity curve will present a exponential decrease form.  In this condition, the adults' cognitive ability is equal to children's level after $600^{th}$ month.

So we can draw the conclusion that the probability of occurrence of the serious cognitive function reduction, such as Alzheimer Disease, is very small. It is due to the stable structure of the human brain's neural network. However, if one has some bad habits that can make sustained damage to his brain neural network, it will cause seriously problem on his cognitive abilities. The sustained damage to the neural network will cause the computational complexity curve became very steep. It also means that the adults' cognitive abilities will decrease more faster. Even in 50 year old, their cognitive abilities will just equal to a child.

This may be one of the reason why Alzheimer Disease appeared. If we can control some factors that can lead to the sustain damage to the neural network early, we may control the sick speed of Alzheimer Disease.

# 4 Conclusion

In this paper, we analyze the neural network's computational complexity of human brain by use the artificial neural network theories. We found that the adults' cognitive abilities decreasing is a very complex process even if we use the very simply full interconnecting neural network. However, the adults cognitive abilities decreasing process is very stable in normal state. If the cognitive decreasing process is along with the psychology function of other organs decreasing, it will not  produce serious problem on adults everyday life. However, if there are some factors, perhaps drugs, smoking, alcohol and so on, that can make sustained damage to the neural network, it will cause serious problems on adults cognitive abilities. It may be one of the reasons that cause the Alzheimer Disease. If we can do more depth research on this model, we may found a way of early prevention of the disease occurred.

# Reference

[1] Cheng, Z., A Cognitive Development Model Based on Neural Network Theories and its Application in Teachers Professional development in Sciencepaper Online(http://www.paper.edu.cn/index.php/default/releasepaper/downPaper/201006-43/1). 2010.

[2] Piaget J, Blanchet A. The grasp of consciousness: Action and concept in the young child[M]. Cambridge, MA: Harvard University Press, 1976.

[3] Cheng, Z. On Adult Leaning Theory Based on Neural Network. Review in Psychology Research. June 2013, Volume 2, Issue 2, PP.27-38